\documentclass[modern]{aastex62}
\hypersetup{linkcolor=red,citecolor=cyan,filecolor=green,urlcolor=magenta}
\newcommand{\gb}{G\"udel-Benz }
\usepackage{upgreek}

\received{XXX}
\revised{XXX}
\accepted{XXX}
\submitjournal{ApJ Letters}
\shorttitle{Peculiar radio--X-ray relationship in active stars}
\shortauthors{Vedantham et al.}

\begin{document}

\title{Peculiar radio--X-ray relationship in active stars}
\correspondingauthor{H. K. Vedantham}
\email{vedantham@astron.nl}

\author[0000-0002-0872-181X]{H. K. Vedantham}
\affil{ASTRON, Netherlands Institute for Radio Astronomy, Oude Hoogeveensedijk 4, Dwingeloo, 7991 PD, The Netherlands}
\affil{Kapteyn Astronomical Institute, University of Groningen, PO Box 72, 97200 AB, Groningen, The Netherlands}

\author[0000-0002-7167-1819]{J. R. Callingham}
\affiliation{Leiden Observatory, Leiden University, PO Box 9513, 2300 RA Leiden, The Netherlands}
\affil{ASTRON, Netherlands Institute for Radio Astronomy, Oude Hoogeveensedijk 4, Dwingeloo, 7991 PD, The Netherlands}
\author[0000-0001-5648-9069]{T. W. Shimwell}
\affiliation{ASTRON, Netherlands Institute for Radio Astronomy, Oude Hoogeveensedijk 4, 7991PD, Dwingeloo, The Netherlands}
\affiliation{Leiden Observatory, Leiden University, PO Box 9513, 2300 RA Leiden, The Netherlands}
\author[0000-0001-9777-9177]{A. O. Benz}
\affil{Institute for Particle Physics and Astrophysics, ETH Zurich, 8093 Z\"urich, Switzerland}
\author[0000-0001-6028-9932]{M. Hajduk}
\affil{Space Radio-Diagnostics Research Centre, University of Warmia and Mazury, ul.Oczapowskiego 2, 10-719 Olsztyn, Poland}
\affil{Department of Astrophysics/IMAPP, Radboud University, P.O. Box 9010, 6500 GL Nijmegen, The Netherlands}
\author[0000-0002-2110-1068]{T. P. Ray}
\affiliation{School of Cosmic Physics, Dublin Institute for Advanced Studies, 
31 Fitzwilliam Place, Dublin, D02 XF86, Ireland}
\author{C. Tasse}
\affil{GEPI \& USN, Observatoire de Paris, Université PSL, CNRS, 5 Place Jules Janssen, 92190 Meudon, France }
\affil{Department of Physics \& Electronics, Rhodes University, PO Box 94, Grahamstown, 6140, South Africa}
\author[0000-0003-2792-1793]{A. Drabent}
\affil{Th\"uringer Landessternwarte, Sternwarte 5, D-07778 Tautenburg, Germany}

\begin{abstract}
The empirical relationship between the non-thermal 5\,GHz radio luminosity and the soft X-ray luminosity of active stellar coronae, canonically called the \gb  relationship \citep{1993ApJ...405L..63G}, has been a cornerstone of stellar radio astronomy as it explicitly ties the radio emission to the coronal heating mechanisms. The relationship extends from microflares on the Sun to the coronae of the most active stars suggesting that active coronae are heated by a flare-like process \citep{1994A&A...285..621B}. The relationship is thought to originate from a consistent partition of the available flare energy into relativistic charges, that emit in the radio-band via the {\em incoherent} gyrosynchrotron mechanism, and heating of the bulk coronal plasma, that emits in the X-ray band via the Bremsstrahlung mechanism. Consequently, {\em coherent} emission from stellar and sub-stellar objects is not expected to adhere to this empirical relationship, as is observed in ultracool dwarf stars and brown dwarfs. Here we report a population of radio-detected chromospherically active stars that surprisingly follows the \gb relationship despite their radio emission being classified as coherent emission by virtue of its high circularly polarised fraction and high brightness temperature. Our results prompt a re-examination of the physics behind the \gb relationship, its implication for the mechanism of coronal heating and particle acceleration in active stars and the phenomenological connection between solar and stellar flares.
\end{abstract}


\section{Introduction}
There exists an empirical quasi-linear relationship, called the \gb relationship,  between the X-ray luminosity ($L_X$\,[${\rm erg}\,{\rm s}^{-1}$]) and the quasi-quiescent, non-thermal 5\,GHz radio spectral luminosity ($L_R\,[{\rm erg}\,{\rm s}^{-1}\,{\rm Hz}^{-1}]$) in chromospherically active stars and binaries: $L_X\approx L_R\times 10^{15.5}$ \citep{1989ApJS...71..905D,1993ApJ...405L..63G,1994A&A...285..621B}. The relationship holds for soft X-ray luminosities ranging from $\sim 10^{29}$ to $10^{32}\,{\rm erg}\,{\rm s}^{-1}$, and the corresponding radio immensities ranging from $10^{22}$ to $10^{28}\,{\rm erg}\,{\rm s}^{-1}$. The soft X-ray emission is due to thermal Bremsstrahlung in the coronal plasma, whereas the radio emission is thought to be due to incoherent gyrosynchrotron emission from relativistic electrons gyrating in the coronal magnetic field. 
The relationship also extends all the way down the energy-scale to microflares from the Sun  that have $L_X\sim 10^{25}\,{\rm erg}\,{\rm s}^{-1}$ \citep{1994A&A...285..621B} suggesting a common flare-like mechanism that deposits energy into the coronae of the Sun and active stars alike.

The \gb relationship is canonically explained in one of two ways, both of which require a consistent fraction of energy in flares to go to accelerating charges to relativistic speeds \citep{1993ApJ...405L..63G,1994A&A...285..621B,2010ARA&A..48..241B}. One way posits that all of the flare energy goes to relativistic charges. These charges lose a tiny fraction of their energy to gyrosynchrotron emission while the bulk of their energy eventually heats up the corona that emits the thermal X-ray emission. The second way is that a small, albeit consistent, fraction of the flare energy goes into relativistic particles while the remainder directly heats the ambient coronal gas. It is also plausible that a different phenomenon (e.g. electric currents or Alfv\'en waves \citep{2000ApJ...530..999M}) drives the coronal heating and particle acceleration, but the consistent energy partitioning remains necessary to explain the relationship.

Although the coronal heating mechanism is not fully understood, the above explanations are generally accepted \citep{2002ARA&A..40..217G} at a qualitative level. However, at the macro-level, the relationship forces a linkage between physical parameters such as the coronal magnetic field strength, cooling timescale of relativistic charges, the fraction of flare energy that go into relativistic charges and the X-ray luminosity \citep{1993ApJ...405L..63G}. These parameters are not readily accessible by independent measurements which has led to a longstanding debate as to the precise micro-physics of the relationship, its implications for the coronal heating problem, and whether the Sun and active stars share a common flare-like coronal phenomenon \citep{1998ApJ...501..805A,1986LNP...254..271H, 2011ASPC..448..455F}. 

More recently, several detections of mildly circularly polarised quasi-quiescent emission were made in ultracool dwarf stars and brown dwarfs. When plotted against the \gb relationship, these objects consistently proved radio-loud or X-ray dim \citep{2001Natur.410..338B,2002ApJ...572..503B,2012ApJ...746...23M,2014ApJ...785....9W}. One explanation offered for the breakdown of the \gb relationship in the ultra-cool dwarf regime is that the radio emission is not due to the gyrosynchrotron mechanism, but due to the coherent cyclotron maser mechanism that appears only mildly polarised because of depolarisation during propagation in the magnetosphere \citep{2008ApJ...684..644H}. An alternate explanation posits that the quasi-quiescent radio emission is gyrosynchrotron but the energetics of these systems or their magnetospheric structure do not support a stable thermal corona to form which leads to their X-ray under-luminosity \citep{2010ApJ...709..332B,2014ApJ...785....9W}. More recently, \citet{2021MNRAS.507.1979L} have argued that the electron acceleration mechanism in dipole dominated objects (from magnetic A/B stars to gas giant planets) is fundamentally different from the flare acceleration seen in the Sun and active stars and that these difference leads to a breakdown of the radio--X-ray relationship in such objects even though the mildly polarised radio emission may be gyrosynchrotron in origin.

Regardless of the above debate on {\em mildly} polarised emission in ultracool-dwarfs, it is widely accepted that {\em highly} polarised radio emission (several tens of per cent) is the result of a coherent emission mechanism and therefore must not adhere to the \gb relationship. This claim is based on the fact that unlike incoherent gyrosynchrotron emission, coherent emission mechanisms in stellar coronae (and planetary magnetosphere) are ultimately driven by plasma instabilities that are sensitive to the {\em gradient} in the electron momentum distribution\footnote{Strictly speaking this is true in the absence of saturation (See also \S3.4).} and not to the {\rm total} energy of the emitting electrons. Additionally, the efficiency of radio emission for incoherent gyrosynchrotron and coherent mechanisms is vastly different. The two mechanisms are therefore not expected to adhere to the same scaling law with coronal X-ray luminosities. Indeed this is the case for highly polarised radio emission at 144\,MHz from M-dwarfs--- they do not follow the \gb relationship, nor establish a new scaling law \citep{2021NatAs.tmp..196C}.

Here we report that highly polarised emission from RS\,Cvn binaries and other high-activity stars at 144\,MHz unexpectedly follows the \gb relationship, upending decades-long and widely-accepted heuristics commonplace in stellar radio astronomy analyses.
Our results prompt a re-examination of the physics behind the \gb relationship and of the mechanism of radio emission from active stars. We outline a few plausible unified models that can reconcile the historical data with our data and motivate future radio observations that can test these models. 

The rest of the paper is organised as follows. \S 2 is devoted to describing the data and ruling out statistical biases as the cause of the anomalous adherence to the \gb relationship in our sample. In \S 3 we discuss possible resolutions to this anomalous result and end with concluding remarks and outlook in \S 4.

\section{The data and statistical tests}
We are presently conducting a systematic survey of the northern sky for stellar, brown dwarf, and exoplanetary radio emissions \citep[see for e.g. ][]{2021A&A...650L..20D,2021NatAs.tmp..196C,2021A&A...648A..13C,2020ApJ...903L..33V,2020NatAs...4..577V}. We primarily identify our targets in the pipeline-processed data from the LOFAR Two-Metre Sky Survey \citep[LoTSS; ][]{2017A&A...598A.104S,2019A&A...622A...1S,2021A&A...648A...1T} as circularly-polarised sources in the Stokes V component of the survey (V-LoTSS; Callingham et al. in prep). Of these, we have previously presented 14 active binaries \citep{2021A&A...654A..21T}. Here, we report six additional chromospherically active stars also discovered as highly circularly polarised radio sources in LoTSS. These stars have circular fractions between $\approx$50\% and 90\%. Since stars close to our Stokes-I detection threshold will not pass the threshold in Stokes~V, we separately searched for Stokes~I emission astrometrically associated stars in the catalogue of chromospherically active binaries compiled by \citet{2008MNRAS.389.1722E}. This yielded four additional RS\,CVn detections all between 0.6 and 0.8\,mJy. Because the typical noise rms values in the LoTSS images are between 70 and 100\,$\upmu$Jy, any polarisation fraction in these sources of about 60\% or lower would not have crossed the detection threshold of our Stokes-V-only search (Callingham et al., in prep.). We include these stars for completeness, although excluding them from our analysis does not change the results and argument presented here. All of our detections are summarised in Table\,\ref{tab:sample}.

The X-ray flux densities for our active sources were determined by identifying their counterparts in the Second ROSAT all-sky survey (2RXS) source catalogue \citep{2016A&A...588A.103B}. We considered an association as reliable when the proper-motion-corrected \emph{Gaia} Data Release 2 \citep[DR2;][]{2018A&A...616A...1G} position of radio-bright active binaries was within 1$'$ of a 2RXS source position. All of our sources are detected in 2RXS, as opposed to the serendipitous survey catalogues produced by \emph{XMM-Newton} or \emph{Chandra}, providing us with a homogeneous X-ray data set. Furthermore, since the original \gb relationship was established with ROSAT data \citep{1994A&A...285..621B}, we do not need to scale or adjust the X-ray data for different wavelength coverage.

The reported 2RXS count rates for our radio detected active stars are listed in Table\,\ref{tab:sample}. The count rate was converted to a 0.1–2.4\,keV flux via the conversion factor $CF = (5.30\mathrm{HR1} + 8.31) \times 10^{-12}$\,erg\,cm$^{-2}$\,count$^{-1}$, where HR1 is the hardness ratio in the soft ROSAT band \citep{1995ApJ...450..401F}. We note that it is likely that $\approx$20\% of our detections were observed during an X-ray flaring event considering the brevity of the ROSAT observations \citep{2012MNRAS.419.1219P}. Such variable X-ray activity does not preclude investigating the \gb relationship since an X-ray flare from an an RS\,CVn binary rarely more than doubles its X-ray flux \citep{2012MNRAS.419.1219P}. The X-ray variability of active binaries will produce scatter within the \gb relationship, as already observed for the canonical \gb data set \citep{1994A&A...285..621B} with its $\sim$0.6 dex scatter around the best fit \citep{2014ApJ...785....9W}.

\begin{figure}
    \centering
    \includegraphics[width=0.8\linewidth]{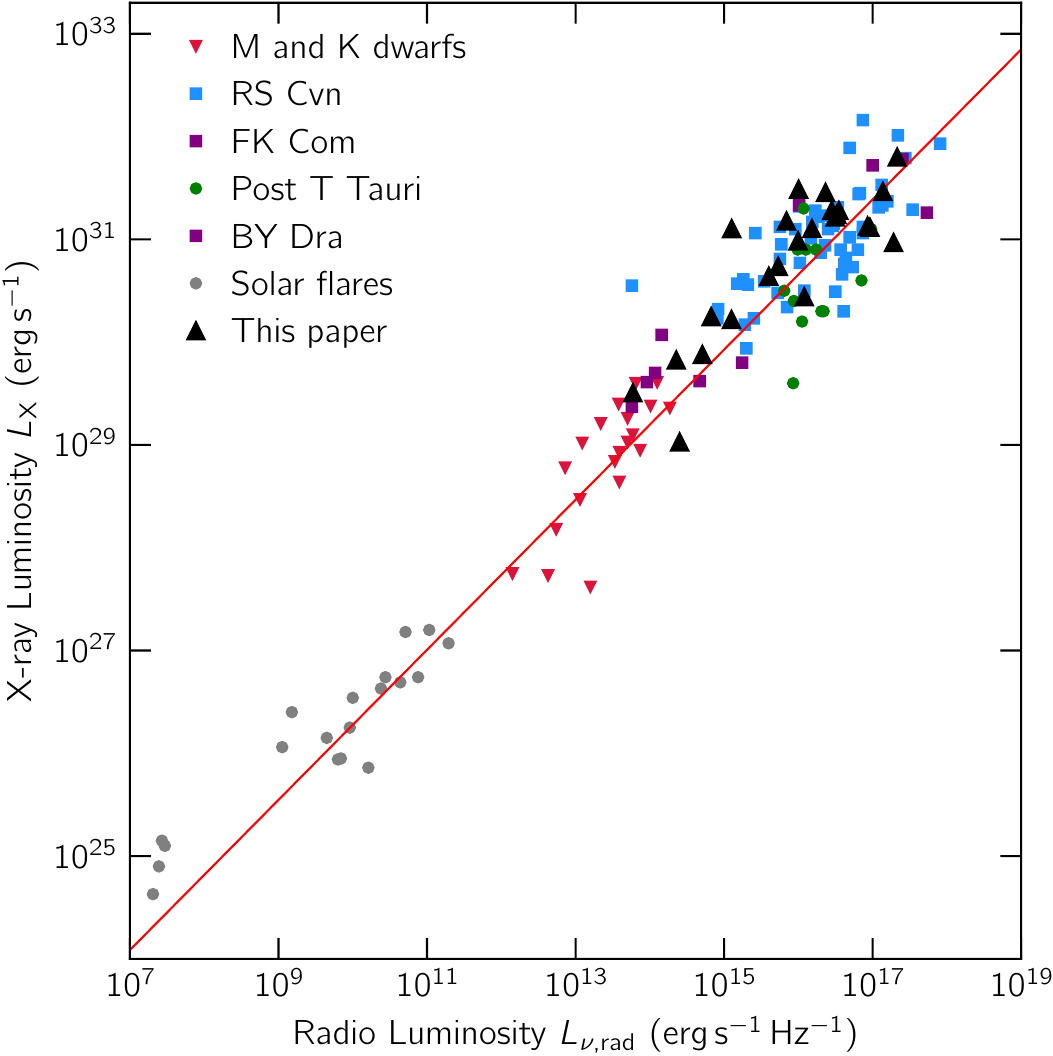}\\
    \includegraphics[width=0.8\linewidth]{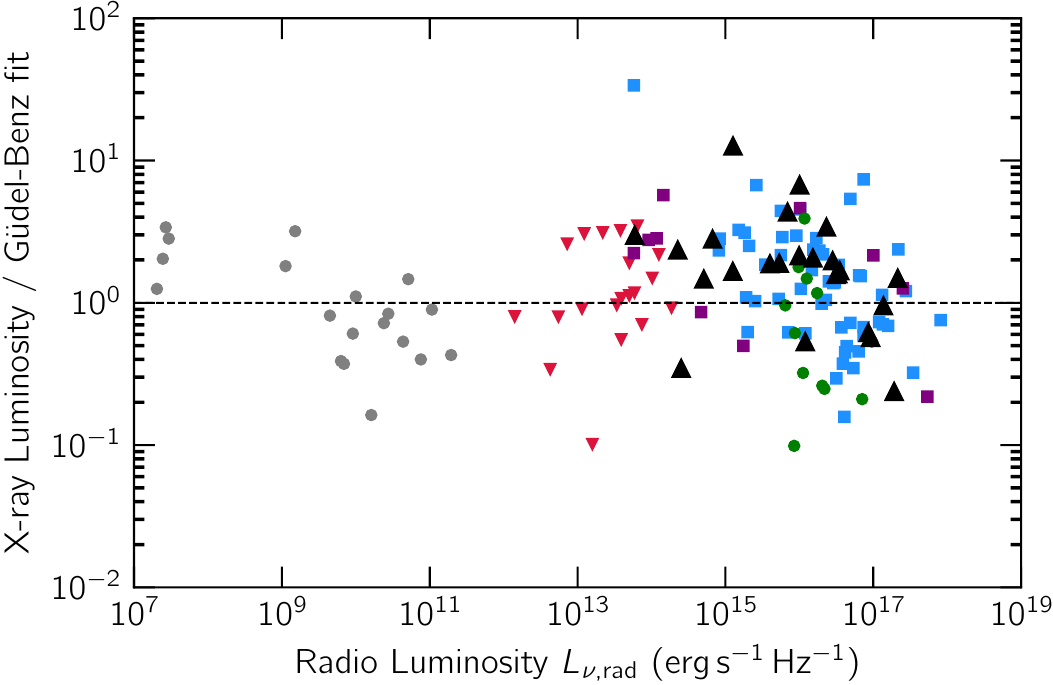}
    \caption{Top-panel: The peculiar adherence of our highly-polarised sample at 144\,MHz (total flux), shown in black triangles, to the \gb relationship shown by the solid red line, $L_{X} = 9.48\times 10^{18} L_{\nu,\,{\rm rad}}^{0.73}$ \citep{2014ApJ...785....9W}. The other points are literature values at 5\,GHz that were initially used to establish the \gb relationship \citep{1994A&A...285..621B}. The X-ray flux is in the 0.1 to 2.4\,keV band. Bottom-panel: Dispersion of the data from the \gb relationship showing that our sample has the same level of scatter as the 5\,GHz data from the literature.}
    \label{fig:gb}
\end{figure}

Figure \ref{fig:gb} shows our sample in the radio-X-ray luminosity plane along with the canonical \gb relationship and the archival sample that led to the discovery of the relationship \citep{1993ApJ...405L..63G,1994A&A...285..621B}. The figure shows a remarkable adherence of our sample to the empirical law. A Kendall-Tau test confirms a monotonic relationship between the radio and X-ray luminosity of our sample with a correlation coefficient of $0.6$ and a p-value for the null hypothesis of $1.3\times10^{-5}$.

The bottom panel of Fig. \ref{fig:gb} shows the scatter of our sample with respect to the \gb relationship. We clearly see that our sample adheres to the \gb relationship as well as the original 5\,GHz sample that was used to establish this empirical relationship.

To ensure that the adherence of our sample to the \gb relationship is neither random happenstance nor a statistical artefact of sample incompleteness inherent in flux-limited surveys (radio in our case), we ran a simple Monte Carlo simulation that produces $1000$ synthetic radio and X-ray data sets under the null hypothesis (i.e. radio and X-ray flux are independent random variables). 
We assumed a space density of X-ray detected active stars that are also radio-emitters of $n=10^{-5}\,{\rm pc}^{-3}$, a distance-horizon of $200\,{\rm pc}$, both consistent with our detections in LoTSS, and a log-normal distribution of X-ray and radio luminosity. The two distributions were assumed to be independent (as consistent with the null hypothesis). 
The simulation horizon was chosen to represent the high Galactic latitude of the fields surveyed so far by our LoTSS data. It also agrees to within a factor of order unity with the farthest radio detection in our population.  The space density of radio emitters was chosen such that the mean number of detection in the Monte-Carlo runs was equal to the number of detections in our sample. 
For the radio luminosity distribution, we assumed a mean of $\left<L_{\rm rad}\right> = 10^{16}\,{\rm erg}\,{\rm s}^{-1}\,{\rm Hz}^{-1}$ and standard deviation of $\sigma_{\rm rad} = 10^1\,{\rm erg}\,{\rm s}^{-1}\,{\rm Hz}^{-1}$, while for the X-ray luminosity we assumed $\left<L_X\right> = 10^{31}\,{\rm erg}\,{\rm s}^{-1}$ and $\sigma_X = 10^{0.75}\,{\rm erg}\,{\rm s}^{-1}$. These values were chosen to approximately span the range of observed radio and X-ray immensities (see Fig. \ref{fig:gb}).

\begin{figure}
    \centering
    \includegraphics[width=0.8\linewidth]{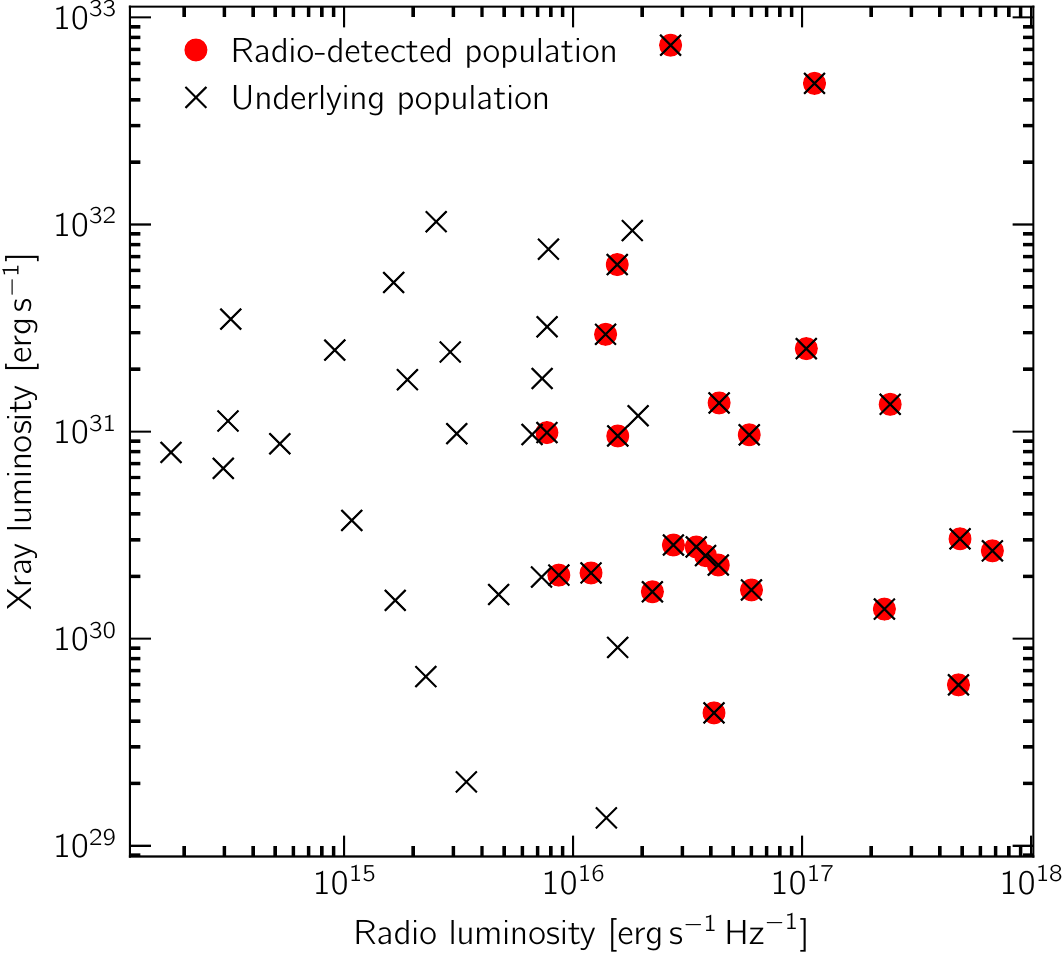}\\
    \includegraphics[width=0.8\linewidth]{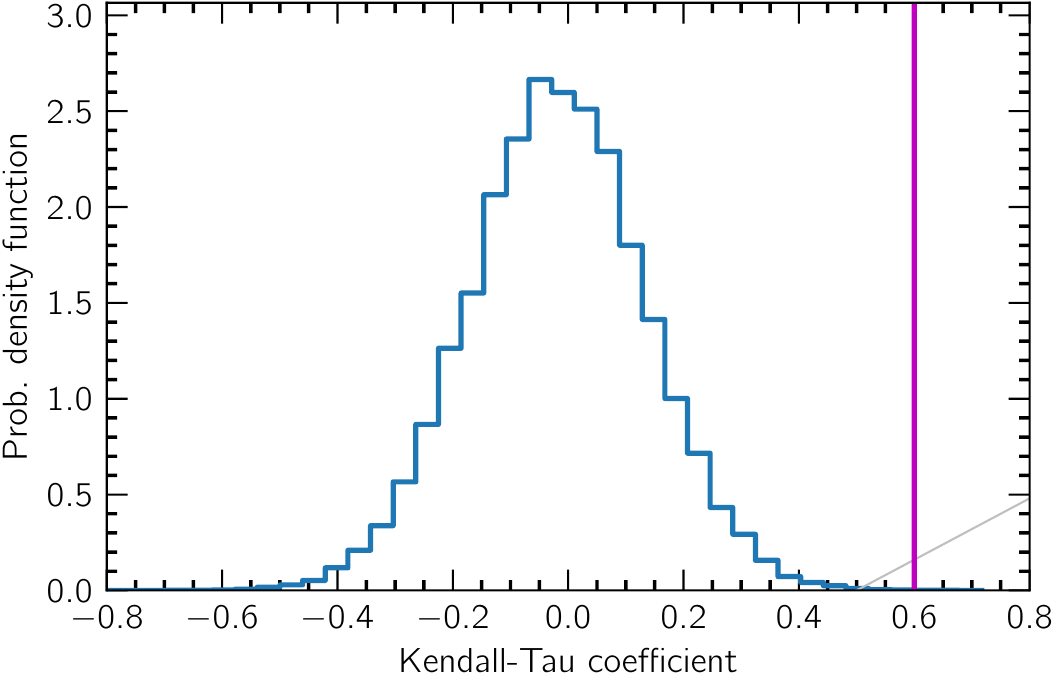}\\
    \caption{Results of the Monte Carlo simulation of radio and X-ray detections under the null hypothesis (i.e. radio and X-ray luminosities are uncorrelated). The top panel shows the sample (black crosses) and the subset that will cross the radio-detection threshold of our survey (red circles), in one realisation. The bottom panel shows the distribution of the Kentall-Tau correlation coefficient between radio and X-ray luminosity of the radio-detected population from $10^5$ realisations (blue step curve) in comparison to the measured coefficient from our sample (magenta line).}
    \label{fig:mcmc}
\end{figure}

The top panel of Figure \ref{fig:mcmc} shows the output of a randomly chosen simulation run in the $L_X,\,L_R$ plane and the bottom panel shows the histogram of the $10^5$ Kendall-Tau rank correlation coefficients from simulations. The simulations reject the null-hypothesis with a p-value of $<10^{-4}$. We, therefore, conclude that our empirical result is not due to chance or a bias due to the volume incompleteness of our radio survey.

All the stars in our sample are highly circularly polarised radio emitters in the metrewave band (Table \ref{tab:sample}) with high brightness temperatures, which suggests that the emission mechanism is a coherent one. \citet{2021A&A...654A..21T} showed that based on the brightness temperature inferred from typical coronal sizes and the emitters' polarised fraction, the emission mechanism is the electron cyclotron maser. On the other hand, the emission mechanism in the centimetre-wave sample that established the \gb relationship is the gyrosynchrotron mechanism. The two radio data sets have also been observed at vastly different radio frequencies and the two emission mechanisms have different emissivities. It is therefore remarkable and puzzling that our sample follows the \gb relationship. It implies that either the inference of the emission mechanism in previous works is wrong or there is a deeper physical reason for why the presumed gyrosynchrotron emission at 5\,GHz emission and the presumed coherent emission at 144\,MHz emission have approximately the same spectral luminosity.

\section{Discussion}

Now that we are faced with an empirically established adherence of highly polarised metre-wave emission to the canonical \gb relationship, we turn our attention to seeking an explanation. In doing so, our aim is not to construct a detailed model of the coronal parameters and their spatial structure but rather to argue from first principles and chart out plausible paths to a resolution.

We will first posit that both the centimetre-wave and metre-wave emission must originate from the same emission mechanism and that the mechanism generates a flat spectrum. This is motivated by the fact that the three possible emission mechanisms which are gyrosynchrotron, plasma, and cyclotron maser all have remarkably different efficiencies and will require contrived scenarios to achieve the observed flux density parity. With this assumption, we have three choices for the common emission mechanism. 

\subsection{Gyrosynchrotron} The centimetre-wave observations have previously been widely interpreted as gyrosynchrptron\footnote{We will use the usual convention and refer to Magnetobremsstrahlung as gyrosynchrotron when emission primarily appears at moderate harmonics between $\approx 10$ to $\approx 100$, and as synchrotron emission in the ultra-relativistic limit.} \citep{2010ARA&A..48..241B}, so we only need to focus our attention on interpreting the metre-wave emission. The main challenges to be addressed here are the high circularly polarised fraction and high brightness temperature at 144\,MHz. The polarisation fraction of our sample spans from about 40\% to 86\%, whereas, the brightness temperature of our sample, assuming a source to be a disk of radius equal to the photometric stellar radius spans from $T_{\rm b}\approx 10^{11.5}\,{\rm K}$ to $T_{\rm b}\approx 10^{13}\,{\rm K}$. 

Synchrotron emission is only weakly circularly polarised even when dispersive effects of the underlying thermal medium (i.e. the Razin-Tsytovich effect \citep{1964ocr..book.....G}) are taken into consideration \citep{1971Ap&SS..12..172M}. The only known regimes where synchrotron emission generates levels of circular polarisation of $\gtrsim 50\%$ are (a) in the presence of a highly anisotropic pitch angle distribution of emitting charges and (b) emission at low harmonics of the cyclotron frequency by mildly relativistic charges\footnote{Emission at low harmonics from highly relativistic charges is very inefficient.}.
\subsubsection{Low pitch-angle regime}
The regime $\gamma\sin\alpha\lesssim 1$, where $\gamma$ is the characteristic electron Lorentz factor and $\alpha$ is the pitch angle is appealing for generating the observed 144\,MHz properties. This is the `small pitch-angle' regime where the ultra-relativistic approximation commonly made \citep[e.g. chapter 6 of ][]{1986rpa..book.....R} breaks down. We refer the reader to \citet{1970ApJ...162L..37O, 1971ApL.....8...35M, 1973ApJ...183..593E} for a discussion of this regime. In this case, the bulk of the radiation is beamed within a cone oriented along the magnetic field with an opening angle given by $\theta\approx\alpha \gamma^{-1}$ and is highly circularly polarised. The emission in the observer frame is the Doppler-boosted cyclotron frequency, given by $\nu =  2\nu_{\rm B}\gamma / (1+\gamma^2\theta^2)$ in the ultra-relativistic limit, where $\nu_{\rm B}$ is the non-relativistic electron cyclotron frequency. The observed emission therefore appears at a frequency of $
\sim \gamma\nu_{\rm B}$ with a fractional bandwidth of order unity. This may be contrasted with the moderate and high pitch angle regimes from canonical theory which yields weakly circularly polarised broad-band emission centred around a much larger critical frequency of $\gamma^2\sin\alpha\nu_{\rm B}$.

With the above theory, a plausible scenario emerges. The electron acceleration process generates relativistic electrons ($\gamma\lesssim 10$) within a narrow forward cone ($\alpha\lesssim 0.1$). These electrons emit in the LOFAR band in the small pitch-angle regime. Their pitch angle distribution eventually broadens due to collision or other processes leading to mildly polarised emission in the cm-wave band by the gyrosynchrotron mechanism. Coming to the issue of brightness temperature, regardless of the kinetic temperature of the emitting electrons and their combined optical depth, the brightness temperature cannot exceed the inverse Compton limit of $\approx 10^{12}\,{\rm K}$.\footnote{Although, in the small pitch angle case, the inverse Compton losses are significantly lowered compared to synchrotron losses \citep{1966ApJ...146..597W}.} Therefore, stars in our sample at the upper end of the brightness temperature limit of $\approx 10^{13}\,{\rm K}$ necessarily imply rather large source sizes of at least a few stellar radii. Hence in this scenario, the emission must come from the bulk of the corona and not the small coronal loops close to the stellar surface. Specific limits on the source size must however await detailed radiative transfer modelling for the postulated case of a directed beam of electrons injected into a large-scale magnetic trap (such as a dipolar trap) in the presence of a background dispersive plasma. 

\subsubsection{Emission at low harmonics}
In the relativistic limit, the critical frequency of synchrotron emission is $\gamma^2\nu_{\rm B}$. In the low-harmonic emission scenario, we are therefore considering $\gamma\gtrsim 1$ charges. Let us take $\gamma\approx 1.5$ as an estimate and postulate that the 144\,MHz emission is at the second harmonic of the ambient cyclotron frequency. The kinetic temperature of the emitting electrons is then $\approx 10^{9.5}\,{\rm K}$ which is the limiting brightness temperature value for incoherent emission. This immediately forces a source size of $10-50$ times the stellar radius. Second harmonic emission additionally forces a magnetic field at the emitter of about 30\,G. Such a strong field cannot be created in-situ via energy available in the stellar wind, because if we equate the magnetic pressure in the interaction region to the ram pressure of the wind, we obtain a wind-density of $\gtrsim 10^{10}\,{\rm cm}^{-3}$ for a $500\,{\rm km}\,{\rm s}^{-1}$ wind speed. The corresponding plasma frequency is $\approx 1\,{\rm GHz}$ which will preclude escape of emission at 144\,MHz. Therefore, the magnetic field must be carried over to tens of stellar radii from the stellar surface by an expanding admixture of thermal and mildly relativistic plasma. This is a model that is reminiscent of moving type-IV sources on the Sun \citep{1973SoPh...32..491D} although on much larger scales and with greater magnetic field strengths. The persistence of the radio emission also requires the type-IV-like phenomena to be ever-present. 

\subsection{Plasma emission}
Fundamental plasma emission can be highly circularly polarised but the emergent radiation can also be depolarised due to propagation effects \citep{1993ASSL..184.....B}. We, therefore, focus our attention on explaining the relatively flat spectrum between 144\,MHz and 5\,GHz that is demanded by the adherence to the \gb relationship.

For a given level of Langmuir wave turbulence, the brightness temperature of the emergent radiation at the plasma frequency, ignoring induced emission (i.e. spontaneous emission only), is expected to be $T_{\rm b}\propto \nu_{\rm p} h_p$ where $\nu_{\rm p}$ is the plasma frequency and $h_p$ is the density scale height \citep{1983SoPh...88..297Z,2001A&A...374.1072S,2021MNRAS.500.3898V}.  Attaining a flat flux-density spectrum in the Rayleigh-Jeans regime, on the other hand, requires $T_{\rm b}\propto (l\nu)^{-2}$ where $l$ is the transverse length-scale of the emitting region. These can be reconciled if $h_p l \propto \nu^{-3}$. 

It is important to note here that in the stars where broad cm-wave observations are available, the emission extends up to at least 15\,GHz \citep{2003A&A...403..613G}. Ensuring a continuous plasma emission spectrum from 144\,MHz to 15\,GHz requires coronal density variations over four orders of magnitude, and the spatial distribution of density variations must follow $h_p(\nu)l(\nu)\propto \nu^{-3}$ or equivalently $h_p(n)l(n)\propto n^{-1.5}$ where $n$ is the density. We are not aware of any physical coronal model that naturally yields these properties. 

Another issue with plasma emission is the seeming absence of non-linear effects in the cm-wave band. 
The implied radio brightness temperature at 144\,MHz requires very high levels of turbulent energy approaching a fraction $w\sim 10^{-5}$ of the background plasma energy density. Such levels of turbulence are expected to lead to induced emission in the cm-wave band \citep{2001A&A...374.1072S} which will lead to flux-densities that are orders of magnitude larger than what is observed at 5\,GHz. 

For the above reasons, we find the plasma emission to be an unlikely cause of the observed radio emission. 

\subsection{Cyclotron maser instability}
Cyclotron maser instability has a very high growth rate. An individual maser site can saturate on millisecond-timescales releasing a fraction of the particle energy \citep{1982ApJ...259..844M}. Such saturation can, in principle, be used to construe a scaling between the radio and X-ray luminosities. As for the radio spectrum, cyclotron maser is empirically known to generate somewhat flat spectra. For example, Jovian cyclotron maser emission has a flux densities that stays within a factor of order unit between $\sim 1\,{\rm MHz}$ and $\approx 40\,{\rm MHz}$ \citep{1998JGR...10320159Z}. The drawback of the cyclotron maser interpretation, however, is that in RS Cvns that have been observed out to 15\,GHz, all show a continuum spectrum that is rather flat \citep{2003A&A...403..613G}. A cyclotron frequency of 15\,GHz implies a field strength of 5.35\,kG which has not yet been observed in Zeeman splitting detections of such stars \citep{1992A&A...265..669D, 2021A&A...650A.197H, 2018MNRAS.481.5163J}.

\subsection{Two mechanisms}
It could also be that the 5--15\,GHz emission and the 144\,MHz emission are due to two different emission mechanisms. If so then it is remarkable that their spectral luminosities must be the same.
To appreciate this, consider a mildly relativistic charge that traverses the magnetic trap of length $L$ over a timescale $L/c$. Cyclotron maser radiation is thought to liberate about 1\% of the particle kinetic energy which means that the energy loss rate due to cyclotron maser radiation for the charge is on average about $0.01 \gamma m_ec^2/(L/c) \approx 10^{-8.5}\gamma \, {\rm erg}\,{\rm s}^{-1}$. The loss rate due to synchrotron emission on the other hand is $\approx 10^{-15}\gamma^2B^2\,{\rm erg}\,{\rm s}^{-1}$, where $B$ is in Gauss. Assuming that cyclotron maser radiation is beamed into a solid angle of about 1\,sr, and taking the radio luminosity to be $\nu L_\nu$, we find that flux density parity between 5\,GHz and 144\,MHz requires $\gamma B^2\approx 10^9$. However, for the cm-wave synchrotron emission to appear at a frequency of 5\,GHz requires $\gamma^2 B \approx 1800$. These two conditions cannot be simultaneously true for reasonable values of $\gamma$ and $B$. We, therefore, conclude that the same charges cannot provide flux-density parity between the 144\,MHz and 5\,GHz emission in this scenario because cyclotron maser instability is simply too efficient.  

Next consider the remaining hypothesis: the cm-wave emission is gyrosynchrotron in origin but the metre-wave emission is due to the plasma mechanism originally powered by the beam instability initiated by the relativistic charges that eventually emit gyrosynchrotron radiation. This scenario is significantly complicated because the plasma emission mechanism is a two-step process. In the first step, the an electron beam generates Langmuir wave turbulence and in the second step, the Langmuir waves undergo non-linear scattering to produce transverse electromagnetic waves. The bulk of the Langmuir wave energy is likely thermalized due to non-linear instabilities (modulational instability for e.g.). Therefore making a feasibility argument based on energetics by appealing to first principles is rather difficult and we do not attempt it here. We can however state that based on the analysis of \citet{2021A&A...654A..21T,2021MNRAS.500.3898V} this hypothesis requires the Langmuir-wave turbulence to approach the strong turbulence limit ($w\sim 10^{-5}$) and for the 144\,MHz emission region to be several stellar radii in size.

\section{Summary \& Outlook}
Long-standing wisdom in the field of stellar radio astronomy states that highly polarised emission is coherent in nature and therefore must not follow the \gb relationship. 
However, here we have demonstrated that the highly circularly polarised 144\,MHz emission from active coronae unexpectedly follows the \gb relationship which was established using the 5\,GHz (presumed) incoherent gyrosynchrotron emission.
Our results, therefore, prompt a return to the drawing board to re-examine the physics behind the \gb relationship and the thus-far presumed mechanism of radio emission from stellar coronae. 
We have taken the first steps in doing so here. We find that if the 144\,MHz and 5\,GHz emission are assumed to be from the same mechanism, then incoherent synchrotron emission from electrons with high pitch-angle anisotropy and incoherent gyroresonance (low-harmonic synchrotron) emission from expanding plasma blobs are the most fruitful avenue to explore, whereas coherent plasma emission and cyclotron maser instability are somewhat less attractive. The incoherent mechanisms are also attractive because coherent emission is inherently highly variable whereas our sample shows a small scatter around the \gb relationship. We find the expanding plasma blobs model to be particularly attractive as it is phenomenologically similar to moving type-IV emission from the Sun \citep{1973SoPh...32..491D}.

If on the other hand, the emission in the two radio bands is due to different mechanisms, then it is remarkable that coronal parameters in our population must conspire to yield a flux-density parity between the two radio bands. In this case, we show that the observed emission cannot be due to cyclotron maser emission at 144\,MHz and gyrosynchrotron emission at 5\,GHz from the same population of charges. The case of plasma emission at 144\,MHz is difficult to readily constrain with available data due to the inherent two-step complexity of the plasma emission mechanism. 

We end by noting that quasi contemporaneous radio observations in the intermediate frequency range ($\approx 200$\,MHz to $\approx 5$\,GHz) are now needed to address the dilemma created by our unusual results. For instance, if the 144\,MHz and 5\,GHz radio emissions originate from different mechanisms then we expect to see two components in the broadband radio spectrum and/or its polarisation properties. Alternatively, if emission in both radio bands is due to the gyrosynchrotron and/or gyroresonance mechanism then we expect a relatively flat spectrum with a predictable smooth change in the polarisation properties across frequency. \\

{\em Acknowledgements}: Software: \texttt{python3, numnpy, scipy, matplotlib}, Telescopes: LOFAR, Services: Vizier, NASA’s Astrophysics Data System. 
We thank Prof. Manuel G\"udel for sharing his published data on radio and X-ray flux densities. We thank Prof. Zarka for commenting on the manuscript.
HKV thanks Prof. Sterl Phinney for discussions. 
JRC thanks the Nederlandse Organisatie voor Wetenschappelijk Onderzoek (NWO) for support via the Talent Programme Veni grant. MH acknowledges the MSHE for granting funds for the Polish contribution to the International LOFAR Telescope (MSHE decision no. DIR/WK/2016/2017/05-1) and for maintenance of the LOFAR PL-612 Baldy (MSHE decision no. 59/E-383/SPUB/SP/2019.1) and the Polish National Agency for Academic Exchange (NAWA) within the Bekker programme under grant No PPN/BEK/2019/1/00431.
TPR acknowledges support from the ERC under grant number 743029 (EASY)
AD acknowledges support by the BMBF Verbundforschung under the grant 05A20STA.
This paper is based on data obtained with the International LOFAR Telescope as part of the LoTSS survey. LOFAR is the Low Frequency Array designed and constructed by ASTRON. It has observing, data processing, and data storage facilities in several countries, that are owned by various parties (each with their own funding sources), and that are collectively operated by the ILT foundation under a joint scientific policy. The ILT resources have benefitted from the following recent major funding sources: CNRS-INSU, Observatoire de Paris and Universit\'{e} d'Orl\'{e}ans, France; BMBF, MIWF-NRW, MPG, Germany; Science Foundation Ireland (SFI), Department of Business, Enterprise and Innovation (DBEI), Ireland; NWO, The Netherlands; The Science and Technology Facilities Council, UK. This research made use of the Dutch national e-infrastructure with the support of the SURF Cooperative (e-infra 180169) and the LOFAR e-infra group. The J\"{u}lich LOFAR Long Term Archive and the German LOFAR network are both coordinated and operated by the J\"{u}lich Supercomputing Centre (JSC), and computing resources on the supercomputer JUWELS at JSC were provided by the Gauss Centre for Supercomputing e.V. (grant CHTB00) through the John von Neumann Institute for Computing (NIC). This research made use of the University of Hertfordshire high-performance computing facility and the LOFAR-UK computing facility located at the University of Hertfordshire and supported by STFC [ST/P000096/1], and of the Italian LOFAR IT computing infrastructure supported and operated by INAF, and by the Physics Department of Turin University (under an agreement with Consorzio Interuniversitario per la Fisica Spaziale) at the C3S Supercomputing Centre, Italy. 

\begin{table}
\begin{tabular}{p{2cm}p{1.5cm}p{2cm}p{2cm}p{1.5cm}p{1.5cm}p{1.5cm}p{1.5cm}}
\hline
\hline
Common name & Type & $S_I$ (mJy) & $S_V$ (mJy) & X-ray lum. (count s$^{-1}$) & Parallax (mas) & Pol. frac (per cent) & X-ray lum (erg\,s$^{-1}$) \\ \hline 

        Sig CrB& RS Cvn& 7.53$\pm$0.28 & -5.82$\pm$0.18 & 9.46 & 47.44 & $-77\pm 3$ & 4.4\\
        BQ CVn & RS Cvn& 2.24$\pm$0.22 & -1.78$\pm$0.15  &0.43  & 5.5982 & $-79\pm 10$ & 13.5\\
        FG UMa & RS Cvn& 0.62$\pm$0.16 & 0.49$\pm$0.10   &0.37 & 4.7389  & $79\pm 26$ & 17.4\\
        BF Lyn & RS Cvn& 1.91$\pm$0.29 & 1.46$\pm$0.16  &3.04  & 42.6303  & $76\pm 14$ & 1.7\\
        DM UMa & RS Cvn& 3.20$\pm$0.26 & -1.72$\pm$0.16 &0.94  & 5.2803 & $-53\pm 6$ & 29.3\\
        EV Dra & RS Cvn& 3.03$\pm$0.28 & -2.61$\pm$0.20 &0.91  & 17.3764 & $-86\pm 10$ & 2.8\\
        DG CVn & RS Cvn& 0.63$\pm$0.11 & -0.57$\pm$0.06 &  0.43& 54.6875 & $-90\pm 18$ & 0.1 \\
        EZ Peg & RS Cvn& 0.87$\pm$0.20 & 0.75$\pm$0.09  &0.67  & 6.0817 & $86\pm 22$ & 19.1\\
        BH CVn & RS Cvn& 2.10$\pm$0.40 & 1.11$\pm$0.11  &2.68  & 21.6693 &  $52\pm 11$ & 5.5 \\
        WW Dra & RS Cvn& 0.54$\pm$0.13 & -0.38$\pm$0.08   &0.51& 6.5160 &  $-70\pm 22$ & 12.7 \\
        YY Gem & RS Cvn& 1.87$\pm$0.26 & -0.87$\pm$0.09   &3.72 & 66.2323 &   $-46\pm 8$ & 0.8\\
        II Peg & RS Cvn& 3.74$\pm$0.37 & 2.61$\pm$0.21   &10.77& 25.4046 & $69\pm 8$ & 15.1 \\
        BD+33 4462& RS Cvn& 1.62$\pm$0.42 & -0.86$\pm$0.20 & 0.2 & 4.5922 & $-53\pm 18$ & 13.1\\
        FG Cam & RS Cvn& 1.83$\pm$0.23 & -1.16$\pm$0.18  &  0.46& 3.1970 &   $-63\pm 12$ & 63.3 \\
        BD+42 2437& Ro. Var& 0.50$\pm$0.13 & -0.37$\pm$0.09  & 0.31  & 4.3684 &  $-74\pm 26$ & 16.5\\
        FK Com & FK Com& 3.40$\pm$0.31 & 2.56$\pm$0.21  & 0.21  & 4.6102 & $75\pm 9$ & 9.3\\
        OU And& Ro Var& 1.51$\pm$0.28 & 1.26$\pm$0.24 &0.69 & 7.1692 & $83\pm 22$ & 19.1\\
        ksi UMa & RS CVn & 0.65$\pm$0.12 & -0.40$\pm$0.12 & 4.92& 114.4867 &  $-61\pm 21$ & 0.3  \\ 
        FI Cnc& FK Com& 0.83$\pm$0.12& 0.69$\pm$0.08 & 0.82 & 10.0082 & $83\pm 15$ & 9.7 \\
        FF UMa& RS Cvn& 1.48$\pm$0.24 & -1.10$\pm$0.13 & 2.20  & 8.7324 &  $-74\pm 14$ & 28.6\\
        44 Boo& W UMa& 1.21$\pm$0.31 & -0.46$\pm$0.12 & 4.73 & 79.95 &  $-38\pm 13$ & 0.7\\ \hline 
        V835 Her& RS Cvn& 0.58 $\pm$ 0.10 & & 1.89  & 32.1839 & & 1.8 \\
        Sig Gem& RS Cvn& 0.72$\pm$ 0.14 & & 8.15&  26.08 & & 12.7\\
        IM Peg& RS Cvn& 0.85$\pm$0.17 & & 2.86& 10.0496 & & 30.8\\
    \end{tabular}
    \caption{List of detected stars in LoTSS DR2 whose radio and X-ray luminosities are plotted in Fig. \ref{fig:gb}. $S_{I}$ and $S_{V}$ correspond to the 144\,MHz total and circularly-polarised flux density, respectively. The X-ray count rate is for the $0.1-2.4$\,keV band from 2RXS \citep{2016A&A...588A.103B}. The parallax measurements are mostly from Gaia DR2 \citep{2018A&A...616A...1G}. For the brightest systems (e.g. ksi\,UMa and 44 Boo), the parallax measurements come from the updated \emph{Hipparcos} catalogue \citep{2007A&A...474..653V}. The error on the polarised fraction is computed by Taylor expanding the quotient ($V/I$) to first order and assuming that the Stokes-V and Stoker-I noise are independent \citep[see Appendix of ][]{2016MNRAS.458.3099V}.
    The horizontal line separates systems with  circularly-polarised counterparts from systems with only detections in total intensity, which are located below the line. Source types are taken from \citet[][and references therein]{2021A&A...654A..21T} and from the Simbad archive server.``Ro. Var'' implies ``rotational variable'' as the exact variable class for BD+42\,2437 is unknown in the literature. Our Stokes V handedness definition is identical to that used by \citet{2021NatAs.tmp..196C}, namely left-handed minus right-handed.}
    \label{tab:sample}
    \end{table}

\end{document}